\documentstyle[prl,aps]{revtex}
\input epsf
\def\NO{\nonumber}
\begin{document}
\title{Monte Carlo Minimization of the Higgs-Landau Potentials}
\author{Jai Sam Kim,$^1$ J.C. Tol\'edano,$^2$ and P. Tol\'edano$^3$\\
{\it 
$^1$Dept of Physics, Pohang University of Science and Technology,
Pohang 790-784, S. Korea\\
$^2$\'Ecole Polytechnique, Rte de Saclay, F-91128, Palaiseau, France\\
$^3$Grp de Physique th\'eorique, Facult\'e des Sciences, 33 rue St-Leu, F-80000,
Amiens, France}}
\date{August 12, 1997}
\maketitle

\begin{abstract}
We propose a Monte Carlo algorithm to search for the boundary points
of the orbit space which is important in determining the symmetry breaking
directions in the Higgs potential and the Landau potential. Our algorithm is
robust, efficient and generally applicable. We apply the method to the Landau 
potential of the $d-$wave abnormal superconductor.

\vskip 0.3cm
\noindent
PACS numbers: 02.70.L, 74.20.De, 11.30.Qc, 12.10.-g
\end{abstract}

\vskip 0.5cm
\section{Introduction}
\vskip 0.3cm

There are many branches of physics where the symmetry breaking is important
yet its analysis is highly non-trivial. The Higgs potential and the 
Coleman-Weinberg potential in the unification theories are difficult to 
minimize except for simple cases. Extremization of the superpotential 
in the supersymmetric unification theories is non-trivial. The Landau potential 
in many branches of condensed matter physics is often times tricky to handle.

The Michel-Radicati conjecture \cite{michel} that a group invariant 
potential has extrema at a few critical strata corresponding to the maximal 
little groups was handy and utilized by many model builders \cite{slan,super}. 
The group theoretical method allows one to locate the extrema from the list
of maximal little groups without solving the coupled third degree extremum
equations. It stirred up considerable research efforts.  Slansky \cite{slan} 
made lengthy tables of little   groups of many representations that could be 
used in the grand unification model building. 
Abud and Sartori \cite{abud} ellaborated the geometrical 
structure of the strata of an irreducible representation. 
Stokes and Hatch \cite{stokes} 
made a complete set of tables for the isotropy groups of the irreducible 
representations of 230 space groups. However, Tol\'edano's \cite{coun} found 
counter-examples to the conjecture. Moreover, the conjecture is less useful
for a reducible representation where strata of maximal little groups
are oftentimes not one-dimensional critical strata.

Some years ago one of the authors \cite{kim1,kim2} devised an efficient method 
for finding symmetry breaking directions in those problems. His method was 
practical and could be used for any representations. The key point in the 
method was the concept of the orbit parameters, which are defined to be 
the dimensionless ratios of ({\it polynomial or non-polynomial}) invariant 
functions to the unique isotropic quadratic invariant. 
For a reducible representation
the ratio is taken over the product of quadratic invariants of each irreducible
representation. Thus an orbit parameter is a kind of angular parameter.
The space of such orbit parameters occupies a confined region, 
which is called the orbit space. Being defined out of invariants each point in 
the orbit space remains unchanged by the symmetry operations and has a definite 
residual symmetry associated with it, its little group, or sometimes called
the isotropy group. Kim showed that the most general quartic Higgs or Landau
potential has the absolute minimum at the most protruding point of the orbit
space, which corresponds most likely to one of the critical strata in the
case of an irreducible representation but not necessarily so in the case of
a reducible representation.  He showed 
that if the Michel-Radicati conjecture is to be valid the orbit space boundary
has to be concave except at the cusps corresponding to the critical strata.
Once the orbit space is constructed then the minimization of the potentials
and the identification of symmetries are easy. One can easily identify the 
minimum point, simply by a look at the orbit space in the two dimensional case.

The advantage of this method is that it finds not only the critical strata
of maximal little groups but also non-critical strata of submaximal little 
groups and the phase diagram can be constructed easily. 
It is powerful if used with the extensive tables made by Slansky \cite{slan}
and Hatch and Stokes \cite{stokes}. It is most useful when the representation 
dimension is large and the orbit space dimension is two or three.

It has been known \cite{johns,mich} that the orbit space is a bounded object 
with a hierarchical structure. The points corresponding to 
the higher symmetries lie hierarchically closer to the surface of the orbit 
space and the interior points correspond to the unique lowest symmetry. 
The points of higher symmetries always form the boundaries of those of lower 
symmetries. The isolated points corresponding to the highest symmetries are
zero dimensional objects and thus the derivatives of the orbit parameters
with respect to the elements of the representation vector, sometimes called
the carrier space vector, are zero. Thus the Michel-Radicati conjecture is 
a natural consequence of the orbit space structure and thus a group invariant
function always has extrema on the critical strata \cite{foot}.
There are additional extrema which depend on the particular choice of the 
invariant potential. In order to determine the absolute minimum one has to 
compare all these local extrema. Thus one has to get the full orbit space 
boundary to locate it. 

Nobody showed that the orbit space boundary of an 
irreducible representation is concave except at the cusps. It is more likely 
that a high dimensional representation has many convex boundary portions as 
the number of available maximal little groups are limited. Moreover, 
the boundaries of an orbit space of a reducible representation have many 
convex boundaries \cite{kim1}.  Nevertheless one can use this group 
theoretical property and systematically search for the little 
groups and the corresponding invariant directions to construct the 
orbit space \cite{slan,stokes}. Several examples were illustrated in
\cite{abud,jari,kim3}. 

One often times deals with the projected subspace of the 
full orbit space. With respect to the subspace the above hierarchy theorem does 
not necessarily hold. When the symmetry group is a non-simple group it is 
awkward to make a list of isotropy groups and the associated invariant 
directions. In this case the above theorem seems to be much less useful. 
Also when the dimension of the representation is large it is non-trivial to 
find the invariant directions corresponding to the little groups.

\vskip 0.5cm
\section{Monte Carlo Search of the Orbit Space}
\vskip 0.3cm
Thus it is necessary to have an alternative method for building the orbit space.
Simple Monte Carlo methods that were tried in the past \cite{kim1,abud} were
not efficient for obtaining the orbit space boundary.
We have devised a more efficient Monte Carlo method for building the orbit 
space. It is applicable to any representation and robust. 
The algorithm uses the von Neumann rejection method \cite{neum} and the 
over-relaxation method. 

Suppose that we are searching for the boundary curves of a two-dimensional
orbit space, $(\theta,\phi)$. 
Let us assume that the orbit parameters $\theta$ and $\phi$ 
are functions of $M$ variables, $x_m$. Then our algorithm is stated as follows:
\begin{quote}
\begin{itemize}
\item[{(1)}]
Initially $N$ points are sampled in a uniformly random way. That is, $N$ sets
of $x_m$ are sampled in a uniformly random way from the interval [-1, 1].
Then one computes the center of gravity, $(\theta_C,\phi_C)$, of these $N$ 
points, $(\theta_i,\phi_i)$.
\item[{(2)}]
For the $N$ points, one computes the angle of the vector,
$(\theta_i-\theta_C,\phi_i-\phi_C)$ and assigns an angular bin. 
Let the total number of bins be $D$.
\item[{(3)}]
One then identifies the outermost points along each of the $D$ discrete
directions. The `outwardness' is defined to be the distance of the
point from the center of gravity.
\item[{(4)}]
Starting from each of these selected points, a random walker with $M-$legs 
makes moves by moving one of its legs, $x_m$. 
If the move increases the distance, it is accepted. 
If not, the random walker steps the leg in the opposite direction with a 
different stride. If the move increases the distance, it is accepted.
Otherwise it is rejected and the old value of $x_m$ is retained.

If the random walker jumps over into another angular bin, its position is
compared with the current outermost point along that direction. 
The more distant of the two
is selected as the new outermost point in that angular bin.
\item[{(5)}]
Step 4 is repeated for all $x_m$.
\item[{(6)}]
Step 4 and 5 are repeated along all the angular directions.
\item[{(7)}]
Step 4, 5 and 6 are repeated a finite times $Rt$ or until no 
further appreciable improvements are made within the allowed number of trials.
\item[{(8)}]
The search of the orbit space boundary is completed.
\end{itemize}
\end{quote}

The circumference of the orbit space is discretized into $D$ bins and
the outermost point in each bin is sought after.
In the random search for the boundary points the random walker of a selected 
bin is encouraged when it moves towards the surface. If
a move $x_m\rightarrow x_m+\delta$ is unsuccessful then another move
$x_m\rightarrow x_m-\omega\cdot\delta$ is tried, where $\omega$ is an
adjustable relaxation parameter.
Even when the random walker strays into another direction its move is not
discarded but compared with the current outermost point. In this way
the random walker finds the outward direction in most attempts. Therefore 
the convergence rate is very fast.

\vskip 0.5cm
\section{Applications}
\vskip 0.3cm
Recently there have been some activities to search for anisotropic
superconductivity \cite{super}. It has been confirmed that a $d$-wave pairing 
actually occurs \cite{tsue}. We  have been investigating a model
of an abnormal superconductor with the high temperature symmetry \cite{tol}, 
$G\equiv O(3)\times U(1)\times T$.
For the spin-0 $d-$wave gap functions the 3-dimensional symmetric traceless 
complex matrix, $\Phi_{ij}$, is assumed for the order parameter.
Under $O(3)$ the indices $i,j$ transform like a three dimensional vector. 
The $U(1)$ transforms $\Phi \rightarrow e^{i\lambda}\Phi$ and 
$\Phi^* \rightarrow e^{-i\lambda}\Phi$. The time reversal symmetry $T$ 
transforms $\Phi \rightarrow \Phi^*$ and vice versa.

Then the most general Landau potential upto the fourth degree in $\Phi$ can
be written as a polynomial of the basic invariant polynomials.
\begin{equation}
V(\Phi)  = 
\alpha I_0 + \beta I_0^2 + \gamma_1 T_1 + \gamma_2 T_2
\label{Landau}
\end{equation}
Each basic invariant polynomial is defined as follows:
\begin{equation}
I_0 = \Phi_{ij}\Phi_{ji}^*
\end{equation}
\begin{equation}
T_1 = \Phi_{ij}\Phi_{jk}^*\Phi_{kl}\Phi_{li}^*, \quad
T_2 = \Phi_{ij}\Phi_{jk}\Phi_{kl}^*\Phi_{li}^*
\end{equation}
where repeated indices imply summation from 1 to 3. It is obvious that
these are invariant under the symmetry group $G$.

The orbit parameters, $\theta$ and $\phi$ are defined as:
\begin{equation}
\theta \equiv {{T_1}\over{I_0^2}},\quad
\phi   \equiv {{T_2}\over{I_0^2}}
\end{equation}
Along the chosen direction $(\theta,\phi)$, the Landau potential has a 
directional minimum value,
\begin{equation}
V_0(\theta,\phi) = 
-{1\over 4}{{\alpha^2}\over{(\beta+\gamma_1\theta+\gamma_2\phi)}}
\end{equation}
and the absolute minimum occurs along a direction where
$(\beta+\gamma_1\theta+\gamma_2\phi)$ has the smallest value.

By choosing an appropriate basis, we can always
have one of the two real matrices in a diagonal form. So
there are actually 7 independent components in $\Phi_{kl}$.
We can define these 7 independent components like:
\begin{equation}
\Phi_{kl}\equiv
\left(
\begin{array}{ccc}
a & 0 & 0 \\
0 & b & 0 \\
0 & 0 & -a-b
\end{array}
\right)
+i\left(
\begin{array}{ccc}
s & u & v \\
u & t & w \\
v & w & -s-t
\end{array}
\right)
\label{redmat}
\end{equation}

We have run the program with the set of parameters: $N=2100$, $Rt=7$,
$D=360$, $\delta=0.125$ and $\omega=0.1$.
The acceptance rate was 0.460. 
The result of the run is as shown in Fig. 1. The orbit space, ($\theta,\phi)$, 
is a triangle with vertices at P1=(1.0,0.0), P2=(0.5,0.5), and P3=(0.\.3,0.\.3).
P1 is associated with the set of variables, $(a=1,b=-1,s=t=0,u=1,v=w=0)$, and
its little group is $D_{4h}\times e^{i\pi} \times T$.
P2 is overlapped by two cusps, ${\rm p_2}=(a=1,b=1,s=t=u=v=w=0)$ corresponding 
to $O^z(2)\times T$ and ${\rm p_4}=(a=1,b=-1,s=t=u=v=w=0)$ corresponding to
$D_{4h}\times e^{i\pi} \times T$.
P3 is associated with $(a=1,b=Q,s=Q,t=1,u=v=w=0)$ where $Q=-2-\sqrt 3$ 
corresponding to the little group $D_{4h}\times e^{i\pi/2} \times T$. 

The little groups of these points are maximal isotropy groups of the high 
temperature symmetry $G$. The spacial part of the isotropy groups of
$P_1$, $P_3$ and $P_4$ are all $D_{4h}$ but their embeddings in $G$ are 
all different.
The elements of these subgroups are listed, in terms of the symbols
used in the {\it International Tables for X-ray Crystallography}, as follows:
\vskip 0.5cm
\begin{equation}
\begin{array}{cl}
H(P_1) = & \{R_1,R_2,R_9,R_{10}\}, \NO \\
 & \{R_3,R_4,R_{11},R_{12}\} \times e^{i\pi}, \NO \\
 & \{R_5,R_6,R_{13},R_{14}\} \times T, \NO \\
 & \{R_7,R_8,R_{15},R_{16}\}\times e^{i\pi} \times T
\end{array}
\end{equation}
\begin{equation}
\hskip -3.2cm
H(P_2) = O^z(2)\times T
\end{equation}
\begin{equation}
\begin{array}{cl}
H(P_3) =
& \{R_1,R_2,R_5,R_6,R_9,R_{10},R_{13},R_{14} \}, \NO \\
& \{R_3,R_4,R_7,R_8,R_{11},R_{12},R_{15},R_{16} \}
\times e^{i\pi/2} \times T
\end{array}
\end{equation}
\begin{equation}
\begin{array}{cl}
H(P_4)  =
& \{R_1,R_2,R_5,R_6,R_9,R_{10},R_{13},R_{14} \} \times T, \NO \\
& \{R_3,R_4,R_7,R_8,R_{11},R_{12},R_{15},R_{16} \} \times e^{i\pi} \times T
\end{array}
\end{equation}

The Landau potential (1) truncated at the fourth degree can have no lower 
symmetry than those of these four points \cite{kim1,kim2}. The degeneracy at P2 
can be lifted if we include the sixth degree terms. (see Fig. 2)

It is interesting that the orbit space has a $convex$ boundary portion in the 
subspace defined by $\phi$ and 
$\xi\equiv(\Phi_{ij}\Phi_{jk}^*\Phi_{kl}\Phi_{lm}\Phi_{mn}^*\Phi_{ni}^*)/I_0^3$.
It is shown in Fig. 2. This implies that if we include the relevant sixth
degree term in the Landau potential non-maximal symmetries may be observed at 
low temperatures depending on the stengths of the expansion coefficients.

\vskip 0.5cm
\section{Discussions}
\vskip 0.3cm
For a $d-$dimensional orbit space, the distance measure is defined in the same
way, i.e., the distance from the random walker's position to 
the center of gravity.
In three dimensions, if the whole range of the two angles, 
polar and azimuthal, are divided into 180 and 360 bins, then the number of bins 
alone is huge, 64800. For even higher dimensions binning the directions is
out of the question. We will need an adaptive Monte Carlo method in this case. 
Namely, we allocate more random walkers in regions of interest. 
However, cusps can be found easily with only a few bins.

There are several parameters that the user selects, the number of angular bins
$D$, the initial number of points $N$, the number of repeats $Rt$, the random 
walker's stride $\delta$, and the relaxation parameter $\omega$.
The optimum choice of these parameters can only be found empirically. 

We tested our method using several Higgs potentials and the Landau potential of
$^3$He. First of all, the outcome
depends on the choice of the center of gravity heavily. So we cannot sensibly
give a performance report in a neat table of numbers. We found that $N$ and $Rt$
must be selected in proportion to the dimension $M$ of the representation.
A large value of $D$ may yield sharper boundary curves but it need not be 
larger than 360. Cusps could be obtained very easily even with very small $D$.

The random walker has to pass many singular curves on its march towards 
the surface. These curves are orbits of submaximal isotropy groups. They are
on the boundary surface in the full orbit space and the Jacobian of the
orbit parameters with respect to $x_m$ vanishes along those curves \cite{kim1}. 
In the projected orbit space parts of the curves are buried inside.
If $\delta$ or $\omega$ is too small the random walker may get stuck at the
curves and may not be able to cross these singular curves. 
So a reasonably large number, $0.25\sim 0.75$, must be chosen as long as 
the rejection rate is not increased too much.

Another point we want to emphasize is that {\it the maximality conjecture is
destined to be broken}. In order to close an $M-$dimensional volume with 
concave boundaries we need at least $(M+1)$ cusps. For example, 3 cusps for 
a two dimensional volume, 4 for three, etc. 
If there are less than $(M+1)$ cusps then portions of the boundaries must be
convex. For example, if we have two cusps for a two dimensional volume then
we can make a banana shape at best instead of a concave triangle.
For a Lie group the dimension of a representation is unlimited whereas
the number of maximal isotropy groups is limited. Thus we have less than
the minimal number of cusps and we will see convex boundaries, which means
the breaking of the maximality conjecture. That is why we see a convex
boundary in Fig. 2.

In conclusion, we have devised a general and robust Monte Carlo algorithm
for finding the symmetry breaking directions. The method can be widely used
both for the Higgs potential and the Landau potential. Applying it to
the Landau theory of the $d-$wave superconductivity, we found that three 
four-fold symmetries are possible at low temperatures, which agrees with
the experiment done on Tl$_2$Ba$_2$CuO$_{6+\delta}$ \cite{tsue}. 

We thank Prof. L. Michel for helpful discussions.
This work was funded by Pohang University of Science \& Technology and
\'Ecole Polytechnique (France).

\vskip 0.5cm
\centerline{FIGURE CAPTIONS}
\vskip 0.5cm
Fig. 1 A random sampling of the orbit space: $(\theta,\phi)$.

Fig. 2 A random sampling of the orbit space: $(\phi,\xi)$.

\vfill\eject
\begin{figure}[h]
\hskip 4.0cm
\epsfxsize=8.5cm
\epsfysize=8.5cm
\epsfbox{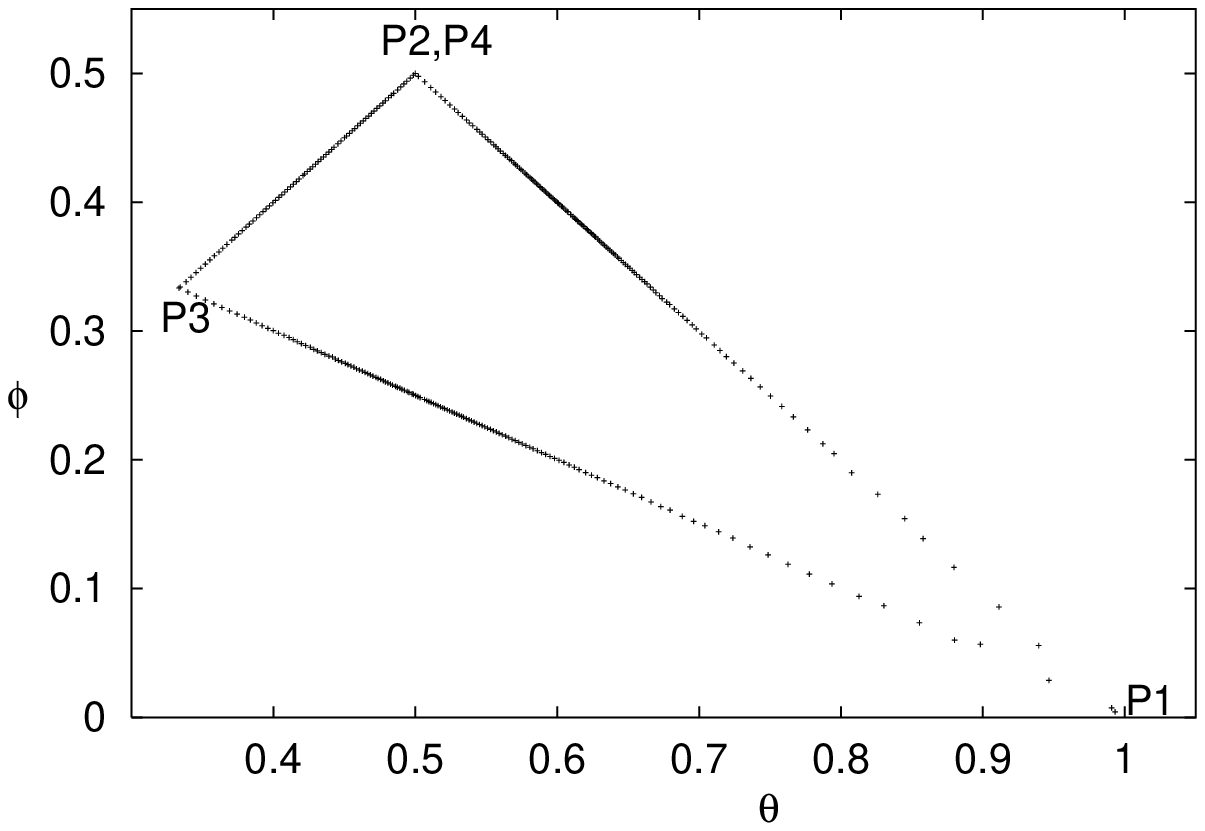}
\caption{A random sampling of the orbit space: $(\theta,\phi)$.}
\end{figure}
\begin{figure}[h]
\hskip 4.0cm
\epsfxsize=8.5cm
\epsfysize=8.5cm
\epsfbox{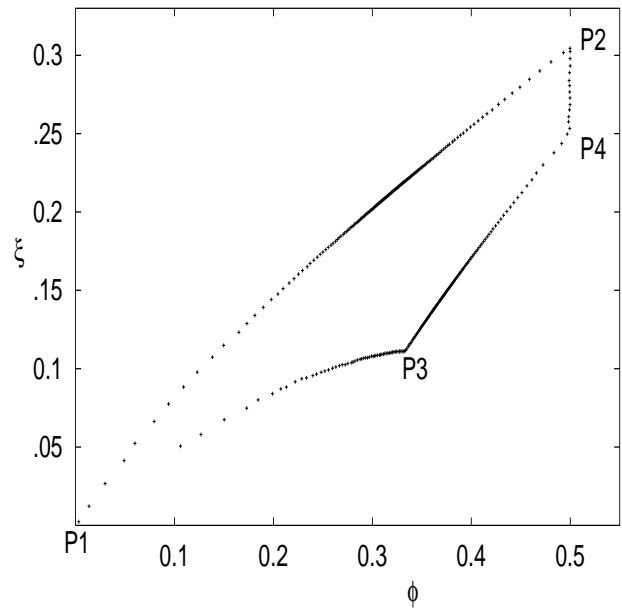}
\caption{A random sampling of the orbit space: $(\phi,\xi)$.}
\end{figure}


\begin{references}
\bibitem{foot}
Recently Michel \cite{mich} obtained a criterion for a critical stratum to be 
a local minimum or maximum in the case of finite groups. He classified
all the extrema of invariant functions defined over the Brillouin zones
for some arithmetic classes in two and three dimensions.
\end{references}
\end{document}